\documentclass{llncs}
\usepackage[english]{babel}

\usepackage{hyperref}
\usepackage{url}
\usepackage{makeidx}

\usepackage[displaymath, mathlines]{lineno}

\usepackage{floatrow}
\usepackage{graphicx}
\usepackage[font=small,labelfont=bf]{caption}
\usepackage[font=small,labelfont=bf]{subcaption}
\usepackage{times}
\usepackage{xspace}
\usepackage{xcolor}

\usepackage{alltt}
\usepackage{amsmath}
\usepackage{amssymb}
\usepackage{tikz}
\usepackage{mathpartir}
\usepackage{listings}
\usepackage{courier}
\usepackage{todonotes}
\usepackage[normalem]{ulem}

\usepackage[all=normal, floats=tight]{savetrees}

\usepackage[title]{appendix}

\usetikzlibrary{shapes}

\begin{document}



\newcommand{\caml}{\textsf{{OCaml}}\xspace}
\newcommand{\C}{\textsf{C}\xspace}

\newcommand{\tool}[1]{\textsf{#1}\xspace}
\newcommand{\framac}{\tool{Frama-C}}
\newcommand{\acsl}{\tool{ACSL}}
\newcommand{\eacsl}{\tool{E-ACSL}}
\newcommand{\compcert}{\tool{CompCert}}
\newcommand{\Value}{\tool{Value}}
\newcommand{\Eva}{\tool{EVA}}
\newcommand{\cfp}{\tool{CfP}}
\newcommand{\tisanalyzer}{\tool{TIS-Analyzer}}
\newcommand{\tisinterpreter}{\tool{TIS-Interpreter}}
\newcommand{\tis}{\tool{TrustInSoft}}
\newcommand{\AES}{\tool{AES}}
\newcommand{\polarssl}{\tool{mbed-TLS}}

\newcommand{\kw}[1]{\texttt{#1}}
\newcommand{\bs}{\ensuremath{\backslash}}
\newcommand{\modif}[1]{{#1}\xspace}

\newcommand{\eg}{\textit{e.g.}\xspace}
\newcommand{\ie}{\textit{i.e.}\xspace}
\newcommand{\etc}{\textit{etc.}\xspace}
\newcommand{\cf}{\textit{cf.}\xspace}

\newcommand{\neginf}{\ensuremath{-\infty}}
\newcommand{\posinf}{\ensuremath{+\infty}}
\newcommand{\zinf}{\ensuremath{\mathbb{Z}_\infty}}


\newcommand{\predone}{P}
\newcommand{\predtwo}{Q}
\newcommand{\defpred}[1]{\kw{defined}(#1)}

\newcommand{\idone}{I}
\newcommand{\idtwo}{H}

\newcommand{\termone}{T}
\newcommand{\termtwo}{U}
\newcommand{\termthree}{V}
\newcommand{\termfour}{W}

\newcommand{\valone}{V}
\newcommand{\valtwo}{W}

\newcommand{\sexpset}{\ensuremath{\mathbb{E}}}
\newcommand{\sexpsetinf}{\ensuremath{\mathbb{E}_\infty}}
\newcommand{\sexpone}{E}
\newcommand{\sexptwo}{D}
\newcommand{\semax}[2]{\kw{max}(#1, #2)}
\newcommand{\semin}[2]{\kw{min}(#1, #2)}

\newcommand{\sidset}{\ensuremath{\mathbb{S}}}
\newcommand{\sidsetpm}{\ensuremath{\sexpset^\pm}}
\newcommand{\sidone}{S}
\newcommand{\sidtwo}{V}

\newcommand{\srangeset}{\ensuremath{\mathbb{R}}}
\newcommand{\srangeone}{R}
\newcommand{\srange}[2]{\left[#1,#2\right]}

\newcommand{\lvtosid}[3]{\ensuremath{#1 \vdash #2 \Leftarrow #3}}
\newcommand{\termtosid}[3]{\ensuremath{#1 \vdash #2 \Rightarrow #3}}

\newcommand{\predtocstr}[3]{\ensuremath{#1 \vdash #2 \Rightarrow #3}}

\newcommand{\bounds}[2]{bounds(#1, #2)}
\newcommand{\tbase}[1]{\ensuremath{\operatorname{base}(#1)}}
\newcommand{\toffset}[1]{\ensuremath{\operatorname{offset}(#1)}}
\newcommand{\trange}[1]{range(#1)}
\newcommand{\proj}[2]{\pi_{#1}(#2)}
\newcommand{\fmin}[2]{min(#1,#2)}
\newcommand{\fmax}[2]{max(#1,#2)}

\newcommand{\memone}{M}
\newcommand{\memtwo}{N}

\newcommand{\lvalone}{L}
\newcommand{\lvaltwo}{\lvalone'}
\newcommand{\memacsone}[1]{\ensuremath{\star\!\! #1}} 
\newcommand{\memacstwo}[1]{\ensuremath{\star #1}} 

\newcommand{\offone}{R}
\newcommand{\offtwo}{S}
\newcommand{\memrange}[2]{\ensuremath{#1\kw{..}#2}}

\newcommand{\intone}{z}
\newcommand{\inttwo}{k}
\newcommand{\intthree}{h}
\newcommand{\intset}{\mathbb{Z}}

\newcommand{\cvarone}{x}
\newcommand{\cvartwo}{y}
\newcommand{\cvarset}{\mathcal{V}}

\newcommand{\pivone}{P}
\newcommand{\pivtwo}{Q}
\newcommand{\pivthree}{R}
\newcommand{\pivset}{\mathcal{P}}
\newcommand{\pivsetof}[1]{\pivset(#1)}

\newcommand{\inter}[2]{ival(#1, #2)} 
\newcommand{\neutral}[1]{neutral\_ival(#1)} 
\newcommand{\ival}[2]{\left[#1;\, #2\right]}
\newcommand{\ivalempty}{\left[-\right]}
\newcommand{\ivalone}{I}
\newcommand{\ivaltwo}{J}

\newcommand{\valid}[1]{\kw{{\bs{}valid}}(#1)}
\newcommand{\rvalid}[1]{\kw{{\bs{}valid\_read}(#1)}}
\newcommand{\init}[1]{\kw{{\bs{}initialized}(#1)}}

\newcommand{\test}[1]{\kw{RTC}(#1)}
\newcommand{\undef}{\kw{UNDEFINED}}
\newcommand{\nalias}[1]{\kw{NEQ}(#1)}
\newcommand{\rtestone}{X} 
\newcommand{\cstrone}{C} 
\newcommand{\intcstrone}{D} 

\newcommand{\env}{\Gamma}
\newcommand{\envof}[1]{\env(#1)}
\newcommand{\envupdate}[2]{\env \left[#1 \mapsto #2\right]}
\newcommand{\id}{id}
\newcommand{\vid}[1]{id_{#1}}

\newcommand{\loc}{\ell}
\newcommand{\range}[2]{\left(#1\kw{..}#2\right)}
\newcommand{\locof}[1]{\mathcal{L}oc(#1)}
\newcommand{\mem}{\Sigma}
\newcommand{\memupone}[3]{#1 \left[#2 \leftarrow #3\right]}
\newcommand{\memadd}[3]{#1 \cup \left\{#2 \mapsto #3\right\}}

\newcommand{\cmp}{\kw{cop}}
\newcommand{\bop}{\kw{bop}}

\newcommand{\solone}{d}
\newcommand{\soltwo}{e}

\newcommand{\type}[1]{type(#1)}
\newcommand{\typing}[2]{#1~:~#2}

\lstdefinelanguage{ACSL}{%
  morekeywords={assert,assigns,assumes,axiom,axiomatic,behavior,behaviors,
    boolean,breaks,complete,continues,decreases,disjoint,ensures,
    exit_behavior,ghost,global,inductive,integer,invariant,lemma,logic,loop,
    model,predicate,reads,real,requires,returns,sizeof,strong,struct,terminates,
    type,union,variant},
  keywordsprefix={\\},
  alsoletter={\\},
  morecomment=[l]{//}
}

\lstdefinestyle{c}{language={[ANSI]C},%
  alsolanguage=ACSL,%
  moredelim={*[l]{//}},%
  moredelim={*[s]{/*}{*/}},%
  moredelim={**[s]{/*@}{*/}},%
  deletecomment={[s]{/*}{*/}},
  moredelim={*[l]{//@}},%
}

\lstset{style=c,
	basicstyle=\scriptsize\ttfamily,
	numberstyle=\tiny,
        keywordstyle=\bfseries,
	numbers=left,
	stepnumber=1,
	numbersep=5pt,
        tab=\rightarrowfill,
}

\frontmatter
\pagestyle{headings}  
\addtocmark{BOH} 

\title{Context Generation from Formal Specifications\\ for C Analysis Tools}
\titlerunning{[Subtitle, if any]}

\author{Michele Alberti\inst{1}%
\thanks{This work was done when the first author was at CEA LIST, Software
  Reliability and Security Laboratory.}
 \and Julien Signoles\inst{2}}
\institute{%
TrustInSoft, Paris, France\\
\email{michele.alberti@trust-in-soft.com}
\and
CEA LIST, Software Reliability and Security Laboratory \\
F-91191 Gif-sur-Yvette Cedex, France \\
\email{julien.signoles@cea.fr}
}

\maketitle

\setlength{\intextsep}{10pt} 
\setlength{\textfloatsep}{10pt} 
\setlength{\abovecaptionskip}{5pt}
\setlength{\belowcaptionskip}{5pt}

\vspace{-4mm} 
\begin{abstract}
Analysis tools like abstract interpreters, symbolic execution tools and testing
tools usually require a proper context to give useful results when analyzing a
particular function. Such a context initializes the function parameters and
global variables to comply with function requirements. However it may be
error-prone to write it by hand: the handwritten context might contain bugs or
not match the intended specification. A more robust approach is to specify the
context in a dedicated specification language, and hold the analysis tools to
support it properly. This may mean to put significant development efforts for
enhancing the tools, something that is often not feasible if ever possible.

This paper presents a way to systematically generate such a context from a
formal specification of a \C function. This is applied to a subset of the \acsl
specification language in order to generate suitable contexts for the abstract
interpretation-based value analysis plug-ins of \framac, a framework for
analysis of code written in \C. The idea here presented has been
implemented in a new \framac plug-in which is currently in use in an operational
industrial setting.

\end{abstract}

\vspace{-6mm} 
\section{Introduction}
\label{sec:introduction}
Code analysis tools are nowadays effective enough to be able to provide suitable
results on real-world code. Nevertheless several of these tools including
abstract interpreters, symbolic execution tools, and testing tools must analyze
the whole application from the program entry point (the \textit{main} function);
or else either they just cannot be executed, or they provide too imprecise
results. Unfortunately such an entry point does not necessarily exist,
particularly when analyzing libraries.

In such a case, the verification engineer must manually write the context of the
analyzed function $f$ as a main function which initializes the parameters of $f$
as well as the necessary global variables. This mandatory initialization step
must enforce the function requirements and may restrict the possible input
values for the sake of memory footprint and time efficiency of the
analysis. This approach is however error-prone: additionally to usual pitfalls
of software development (\eg bugs, code maintenance, \etc), the handwritten
context may not match the function requirements, or be over
restrictive. Moreover this kind of shortcomings may be difficult to detect due
to the fact that the context is not explicitly the verification objective.

A valid and more robust alternative is to specify such a context in a dedicated
specification language, and make the analysis tools handle it properly. This is
often an arduous approach as the support for a particular specification language
feature may entail a significant development process, something that is often
not feasible if ever possible. Also, it requires to do so for every tool.

This paper presents a way to systematically generate an analysis context from a
formal specification of a \C function. The function requirements as well as the
additional restrictions over the input domains are expressed as function
preconditions in the \textsf{ANSI/ISO} \C Specification Language (in short,
\acsl)~\cite{acsl}. This specification $\mathcal{S}$ is interpreted as a
constraint system, simplified as much as possible, then converted into a \C code
$\mathcal{C}$ which exactly implements the specification $\mathcal{S}$. Indeed
not only every possible execution of $\mathcal{C}$ satisfies $\mathcal{S}$ but
conversely, there is an execution of $\mathcal{C}$ for every possible input
satisfying the constraints expressed by $\mathcal{S}$. We present the
formalization of this idea for an expressive subset of \acsl including standard
logic operators, integer arithmetic, arrays and pointers, pointer arithmetic,
and built-in predicates for the validity and initialization properties of memory
location ranges.

We also provide implementation details about our tool, named \cfp for
\emph{Context from Preconditions}, implemented as a \framac plug-in. \framac is
a code analysis framework for code written in \C~\cite{kirchner15fac}. %
Thanks to the aforementioned technique, \cfp generates suitable contexts for two
abstract interpretation-based value analysis tools, namely the the \framac
plug-in \Eva~\cite{blazy17vmcai} and \tisanalyzer~\cite{cuoq17jfla} from the
\tis company. Both tools are actually distinct evolved versions of an older
plug-in called \Value~\cite{canet09scam}. In particular, 
\tis successfully used \cfp on the \polarssl library (also known as
\tool{PolarSSL}), an open source implementation of
SSL/TLS\footnote{\url{https://tls.mbed.org/}}, when building its verification
kit~\cite{polarssl-report}. It is worth noting that \cfp revealed some mistakes
in contexts previously written by hand by expert verification engineers when
comparing its results with these pieces of code.  Also, \cfp generates code as
close as possible to human-written code: it is quite readable and follows code
patterns that experts of these tools manually write.


  \paragraph{Contributions} The contributions of this paper are threefold:
  \textbf{a novel technique to} systematically \textbf{generate an analysis
    context} from a formal specification of a \C function, \textbf{a precise
    formalization} of this technique, and a presentation of \textbf{a tool}
  implementing this technique which is \textbf{used in an operational industrial
    setting}.


  \paragraph{Outline} Section~\ref{sec:overview} presents an overview of our
  technique through a motivating example.  Section~\ref{sec:prec_to_cstr}
  details preconditions to constraints conversion, while
  Section~\ref{sec:cstr_to_code} explains the \C code generation scheme for
  these latter. Section~\ref{sec:evaluation} evaluates our approach and
  Section~\ref{sec:related_work} discusses related
  work. Section~\ref{sec:conclusion} concludes this work by also discussing
  future work.

\vspace{-3mm} 
\section{Overview and Motivating Example}
\label{sec:overview}
We illustrate our approach on context generation through the function
\kw{aes\_crypt\_cbc}, a cryptographic utility implemented by the \polarssl
library. Figure~\ref{fig:contract} shows its prototype and \acsl preconditions
as written by \tis for its verification kit~\cite{polarssl-report}.

\begin{figure}[htb]
\vspace{-2mm} 
\begin{lstlisting}
typedef struct {
  int nr;                     /*  number of rounds  */
  unsigned long *rk;          /*  AES round keys    */
  unsigned long buf[68];      /*  unaligned data    */
} aes_context;

/*@ requires ctx_valid: \valid(ctx);
  @ requires ctx_init: \initialized(ctx->buf + (0 .. 63));
  @ requires ctx_rk: ctx->rk == ctx->buf;
  @ requires ctx_nr: ctx->nr == 14;
  @ requires mode: mode == 0 || mode == 1;
  @ requires length: 16 <= length <= 16672;
  @ requires length_mod: length % 16 == 0;
  @ requires iv_valid: \valid(iv + (0 .. 15));
  @ requires iv_init: \initialized(iv + (0 .. 15));
  @ requires input_valid: \valid_read(input + (0 .. length - 1));
  @ requires input_init: \initialized(input + (0 .. length - 1));
  @ requires output_valid: \valid(output + (0 .. length - 1)); */
int aes_crypt_cbc(aes_context *ctx,int mode,size_t length,unsigned char iv[16],
                  const unsigned char *input,unsigned char *output);
\end{lstlisting}
\vspace{-2mm} 
\caption{\acsl preconditions of the \polarssl function
  \texttt{aes\_crypt\_cbc}.}
\label{fig:contract}
\end{figure}

\vspace{-2mm} 
\paragraph{Specification}

The function \kw{aes\_crypt\_cbc} provides encryption and decryption of a buffer
according to the \AES cryptographic standard and the CBC encryption mode. The
function takes six parameters. The last two are the input and the output
strings.  The parameter \kw{ctx} stores the necessary information to the \AES
substitution-permutation network, in particular the number of rounds and the
round keys defined in a dedicated structure at lines 1--5.  The parameter
\kw{mode} indicates whether the function should encrypt or decrypt the
\kw{input}. The parameter \kw{length} indicates the length of the \kw{input}
string. Finally the parameter \kw{iv} provides an initialization vector for the
\kw{output} of 16 characters (\texttt{unsigned char iv[16]}). This declared
length is actually meaningless for most \C tools because an array typed
parameter is adjusted to have a pointer type~\cite[Section 6.9.1 and also
footnote 79 at page~71]{c99}, but \cfp nevertheless considers it as part of the
specification in order to generate a more precise context.

\acsl annotations are enclosed in \texttt{/*@ ... */} as a special kind of
comments. Therefore they are ignored by any \C compiler. A function precondition
is introduced by the keyword \lstinline+requires+ right before the function
declaration or definition. It must be satisfied at every call site of the given
function. Here the function \texttt{aes\_crypt\_cbc} has 12 precondition
clauses, and the whole function precondition is the conjunction of all of
them. Clauses may be tagged with names, which are logically meaningless but
provide a way to easily refer to and to document specifications. For instance,
the first precondition (line 7) is named \texttt{ctx\_valid} while the second
(line 8) is named \texttt{ctx\_init}.

We now detail the meaning of each precondition clause. All pointers must be
valid, that is properly allocated, and point to a memory block of appropriate
length that the program can safely access either in read-only mode (predicate
\lstinline+\valid_read+), or in read-write mode (predicate \lstinline+\valid+)
. That is the purpose of preconditions \texttt{ctx\_valid}, \texttt{iv\_valid},
\texttt{input\_valid} and \texttt{output\_valid}: \texttt{ctx} must point to a
memory block containing at least a single \texttt{aes\_context} struct,
\texttt{iv} must be able to contain at least 16 unsigned characters (ranging
from 0 to 15), while \texttt{input} and \texttt{output} must be able to contain
at least \texttt{length} unsigned characters (ranging from 0 to
$\texttt{length}-1$).  Memory locations, which are read by the function, must be
properly initialized. That is the purpose of the precondition clauses
\texttt{ctx\_init}, \texttt{iv\_init}, and \texttt{input\_init} which initialize
the first 64 cells of \texttt{ctx->buf} as well as every valid cell of
\texttt{iv} and \texttt{input}. The specification clause \texttt{mode} specifies
that the mode must be either 0 (encryption) or 1 (decryption), while the
specification clause \texttt{length\_mod} specifies that the length should be a
multiple of the block size (\emph{i.e.} 16) as specified in \polarssl. The other
clauses restrict the perimeter of the analysis in order to make it tractable.

The clause \texttt{ctx\_rk} is a standard equality for an \AES context, while
the clause \texttt{ctx\_nr} is true for 256-bit encryption keys. Finally the
clause \texttt{length} aims to restrict the analysis to buffers of size from 16
to 16672 unsigned characters.

\paragraph{Context Generation} 

A naive approach for context generation would consider one precondition clause
after the other and directly implement it in \C code. However, this would not
work, in general, since requirements cannot be treated in any order. In our
running example, for instance, variables \kw{input} and \kw{output} depends on
the variable \kw{length}: the precondition clauses over this latter must be
treated before those over the former, as well as the generated code for these
variables must initialize the latter, first, and the former afterwards, to be
sound. 
To solve such problems, one could first record every dependency among the
left-values involved in the specification, and then proceed to generate \C code
accordingly. An approach based only on a dependency graph is nonetheless
insufficient for those preconditions that need an inference reasoning in order
to be implemented correctly. As an example, treating the precondition
\lstinline{/*@ requires \valid(x+(0..3)) && *(x+4)==1;*/} demands to infer
\kw{x} as an array of 5 elements in order to consider the initialization
\lstinline{x[4] = 1;} correct.

We now give an overview on how we treat context generation by means of the
plug-in \cfp of \framac. On the \kw{aes\_crypt\_cbc} function
contract, \cfp provides the result shown in Figure~\ref{fig:generate} (assuming
that the size of \kw{unsigned long} is 4 bytes\footnote{This kind of
system-dependent information is customizable within \framac.}).

\begin{figure}[htb]
\begin{lstlisting}
int cfp_aes_crypt_cbc(void) {
  unsigned char *cfp_output, *cfp_input;
  unsigned char cfp_iv[16];
  size_t cfp_length;
  aes_context cfp_ctx;
  int cfp_disjunction;
  cfp_length = Frama_C_unsigned_int_interval(16, 16672);
  if (cfp_length % 16 == 0) {
    Frama_C_make_unknown((char *)cfp_ctx.buf,256);
    cfp_ctx.nr = 14;
    cfp_ctx.rk = cfp_ctx.buf;
    Frama_C_make_unknown((char *)cfp_iv,16);
    cfp_input = (unsigned char *)malloc(cfp_length);
    if (cfp_input != (unsigned char *)0) {
      Frama_C_make_unknown((char *)cfp_input, cfp_length);
      cfp_output = (unsigned char *)malloc(cfp_length);
      if (cfp_output != 0) {
        cfp_disjunction = Frama_C_int_interval(0,1);
        if (cfp_disjunction) {
          int cfp_mode;
          cfp_mode = 1;
          aes_crypt_cbc(&cfp_ctx,cfp_mode,cfp_length,cfp_iv,cfp_input,cfp_output);
        }
        else {
          int cfp_mode;
          cfp_mode = 0;
          aes_crypt_cbc(&cfp_ctx,cfp_mode,cfp_length,cfp_iv,cfp_input,cfp_output);
        }
      }
    }
  }
  return 0;
}
\end{lstlisting}
\vspace{-2mm} 
\caption{Slightly simplified version of the code generated by \cfp for the
  specification in Figure~\ref{fig:contract}. Compared to the actual version,
  only a few integer casts have been removed for reasons of brevity.}
\label{fig:generate}
\vspace{-2mm} 
\end{figure}

First note that every execution path ends by a call to the function
\kw{aes\_crypt\_cbc}. Up to these calls, the code initializes the context
variables (prefixed by \kw{cfp}) in order to satisfy the precondition of this
function, while the different paths contribute to cover all the cases of the
specification. The initialization code is generated from sets of constraints
that are first inferred for every left-value involved in the precondition. While
inferring these constraints from the precondition clauses, the implicit
dependencies among left-values are made explicit and recorded in a dependency
graph. This latter is finally visited to guide the code generation process in
order to obtain correct \C code.

Let us start detailing the generated code for both preconditions about
\kw{length} (Figure~\ref{fig:contract}, lines 12--13). First \cfp declares a
variable \kw{cfp\_length} of the same type as \kw{length} (line 4). Then it
initializes it by means of the \framac library function
\kw{Frama\_C\_unsigned\_int\_interval} (line~7). It takes two \kw{unsigned int}
arguments and returns a random value comprised between the two. This allows to
fulfill the former requirement and to guarantee that \framac-based abstract
interpreters will interpret this result with exactly the required
interval. Also, it corresponds to the way that expert engineers would write a
general context for such analyzers. Finally, the requirement \kw{length \% 16 ==
  0} is implemented by the conditional at line~8.

Lines 9--11 implement the preconditions about \kw{ctx}, a pointer to an
\kw{aes\_context}. Instead of allocating such a pointer, the generated code just
declares a local variable \kw{cfp\_ctx} and passes its address to the function
calls. This automatically satisfies the precondition on pointer
validity. Line~9 initializes the 256 first bytes of the structure field
\kw{buf} by using the \framac library
function \kw{Frama\_C\_make\_unknown}. Assuming that the size of \kw{unsigned
  long} is 4 bytes, 256 bytes is the size of 64 values of type \kw{unsigned 
  long}. Again, an expert engineer would also use this library
function. Lines~10 and~11 initialize the fields \kw{ctx->nr} and \kw{ctx->rk}
by single assignments. Here \cfp fulfills the equality requirement
$\texttt{ctx->rk}==\texttt{ctx->buf}$ with respect to \kw{ctx->rk} instead of
\kw{ctx->buf} because the latter already refers to a memory buffer.

The requirements on function arguments \kw{iv}, \kw{input}, and \kw{output} are
implemented by lines 12--17. Let us just point out how \cfp defines the
respective variables: while \kw{ctx\_iv} is as an array of $16$ \kw{unsigned
  char}, \kw{ctx\_input} and \kw{ctx\_output} are just pointers to dynamically
allocated memory buffers. Indeed, while \cfp can infer the exact dimension of
the former from the specification, the dimension of these latter depends on the
value of \kw{ctx\_length}, which is determined only at runtime.

The last part of the generated code (lines 18--29) handles the requirement on
\kw{mode}, which is either 0 or 1. Although the generated conditional may seem
excessive in the case of these particular values, it is nonetheless required in
the general case (for instance, consider the formula \kw{mode == 5 || mode ==
  7}).





\vspace{-3mm} 
\section{Simplifying \acsl Preconditions into State Constraints}
\label{sec:prec_to_cstr}

This section presents a way to systematically reduce a function precondition to
a set of constraints on the function context (\emph{i.e.} function parameters
and global variables).

We first introduce an \acsl-inspired specification language on which we shall
formalize our solution. Then, we define the notion of state constraint as a form
of requirement over a \C left-value, which in turn we generate as \C code for
initializing it. In order to simplify state constraints the most, we make use
of symbolic ranges, originally introduced by Blume and
Eigenmann~\cite{BlumeEigen} for compiler optimization. We finally provide a
system of inference rules that formalizes such a simplification process.
\vspace{-2mm} 

\subsection{Core Specification Language}
\label{sec:acsl_prec_lang}

In this work we shall consider the specification language in
Figure~\ref{fig:syntax_logic}. It is almost a subset of \acsl~\cite{acsl} but
for the predicate \kw{defined}, which subsumes the \acsl predicates
\lstinline+\initialized+ and \lstinline+\valid+ (see below).

\begin{figure*}[htbp]
\vspace{-2mm} 
\[\arraycolsep=4pt
\begin{array}{lrll}
  \textnormal{Predicates} & \predone ::=
  & \termone~\kw{cop}~\termone & \textnormal{term comparison ($\kw{cop}\in\{\equiv,\leq,<,\geq,>\}$)} \\
  & \mid & \defpred{\memone} & \textnormal{$\memone$ is defined}\\
  & \mid & \predone \wedge \predone \mid \predone \vee \predone \mid \neg \predone & \textnormal{logic formula}
  \\[2mm]

  \textnormal{Terms} & \termone ::=
  & \intone & \textnormal{integer constant ($\intone\in\intset$)}\\
  & \mid & \memone & \textnormal{memory value}\\
  & \mid & \termone~\kw{bop}~\termone & \textnormal{arithmetic operation ($\kw{bop}\in\{\kw{+},\kw{-},\times,\kw{/},\kw{\%}\}$)}
  \\[2mm]


\textnormal{Memory Values} & \memone ::= & \lvalone & \textnormal{left-value} \\
  & \mid & \memone~\kw{++}~\termone & \textnormal{single displacement}\\
  & \mid & \memone~\kw{++}~\memrange{\termone}{\termone} & \textnormal{displacement range}
  \\[2mm]

  \textnormal{Left-Values} & \lvalone ::= & \cvarone & \textnormal{\C
    variable} \\
  & \mid & \star \memone & \textnormal{dereference}\\[2mm]

  \textnormal{Types} & \kappa ::= & \iota &
  \textnormal{integer} \\
  & \mid & \kappa \star & \textnormal{pointer}
\end{array}
\]
\vspace{-4mm} 
\caption{Predicates, terms, and types.}\label{fig:syntax_logic}
\end{figure*}

Predicates are logic formul{\ae} defined on top of typed term comparisons and
predicates \kw{defined}. Terms are arithmetic expressions combining integer
constants and memory values by means of the classic arithmetic operators. 
Memory values include left-values, which are \C variables and pointer
dereferences ($\star$), and memory displacements through the operator
($\kw{++}$). 
In particular, $\memone~\kw{++}~\memrange{\termone_1}{\termone_2}$ defines the
set of memory values $\{\memone~\kw{++}~\termone_1, \dots,
\memone~\kw{++}~\termone_2\}$ and may only appear as the outermost construct in
a predicate \kw{defined}.
%
On integers, \kw{defined}($\lvalone$) holds whenever $\lvalone$ is an
initialized left-value. On pointers, \kw{defined}($\memone$) holds whenever
$\memone$ is a properly allocated and initialized memory
region.

\paragraph{Term typing}
Terms of our language are typed. 
A left-value may take either an
integer~($\iota$) or a pointer~($\kappa\star$) type, while memory values are
pointers. We omit the typing rules for terms, which are quite standard
. Let us just specify that memory values of the form $\memone~\kw{++}~\termone$
have pointer type, as well as the recursive occurrence $\memone$, while $T$ must
have integer type. (Memory values
$\memone~\kw{++}~\memrange{\termone}{\termone}$ are typed as set of
pointers~\cite{acsl}.) Since we do not consider any kind of coercion construct,
terms of pointer type cannot appear where integer terms are expected, that is,
they cannot appear in arithmetic expressions. It also follows that term
comparisons only relate terms of the same type.

\paragraph{Term normal forms}
For the sake of concision and simplicity, the remainder of this work assumes
some simplifications to take place on terms in order to consider term normal
forms only. In particular, arithmetic expressions are maximally flattened and
factorized (\eg by means of constant folding techniques,
\etc). 
We will conveniently write single displacements $\memone~\kw{++}~\termone$ as
$\memone~\kw{++}~\memrange{\termone}{\termone}$. We also assume memory values
with displacement ranges to be either of the form
$\cvarone~\kw{++}~\memrange{\termone_1}{\termone_2}$ or
$\star\lvalone~\kw{++}~\memrange{\termone_1}{\termone_2}$. To this end, terms of
the form
$(\lvalone~\kw{++}~\memrange{\termone_1}{\termone_2})~\kw{++}~\memrange{\termone_3}{\termone_4}$
simplify into
$\lvalone~\kw{++}~\memrange{(\termone_1~\kw{+}~\termone_3)}{(\termone_2~\kw{+}~\termone_4)}$.
Finally, memory values $\lvalone~\kw{++}~\memrange{0}{0}$ normalize to
$\lvalone$.

\paragraph{Disjunctive normal forms}

A precondition is a conjunction of predicate clauses, each one given by an \acsl
\lstinline+requires+ (\cf example in Figure~\ref{fig:contract}). As a
preliminary step, we shall rewrite this conjunctive clause into its disjunctive
normal form $\bigvee_i \bigwedge_j \predone_{ij}$, where each $\predone_{ij}$ is
a \emph{predicate literal} (or simply \emph{literal}), that is, a predicate
without nested logic formul{\ae}. A \emph{negative literal} is either of the
form $\neg \defpred{\memone}$ or $\neg (\memone_1 \equiv \memone_2)$, with
$\memone_1,\memone_2$ pointers, as every other negative literal in the input
predicates is translated into a positive literal by applying standard arithmetic
and logical laws. A non-negative literal is called a \emph{positive
  literal}. Most of the rest of this section focuses on positive literals:
  negative 
literals and conjunctive clauses are handled in the very end, while disjunctive
clauses will be considered when discussing code generation in
Section~\ref{sec:cstr_to_code}.

\subsection{State Constraints}
\label{sec:state_constraints}


We are interested in simplifying a predicate literal into a set of constraints
over \C left-values, called \emph{state constraints}. These are meant
to 
indicate the minimal requirements that the resulting \C function context must
implement for satisfying the function precondition. In
Section~\ref{sec:cstr_to_code}, they will be, in turn, converted into \C code.

We intuitively consider a state constraint to represent the domain of definition
of a \C left-value of the resulting function context state. Since such domains
might not be determined in terms of integer constants only, we shall found their
definition on the notion of symbolic ranges~\cite{BlumeEigen}. 
%
As we want to simplify state constraints the most, we 
define them in terms of the symbolic range algebra proposed by Nazar\'e et
al.~\cite{Na14}. Our definitions are nonetheless significantly different, even
though inspired from their work.

\paragraph{Symbolic Expressions}
A \emph{symbolic expression} $\sexpone$ is defined by the following grammar,
where $\intone\in\intset$,
$\kw{bop}\in\{\kw{+},\,\kw{-},\,\times,\,\kw{/},\,\kw{\%}\}$, and \kw{max} and
\kw{min} are, respectively, the largest and the smallest expression
operators. We denote $\sexpset$ the set of symbolic expressions.
\begin{linenomath}
\[
\sexpone ::= \intone~|~\cvarone~|~\memacsone{\sexpone}~|~\sexpone~\kw{bop}~\sexpone~|~\semax{\sexpone}{\sexpone}~|~\semin{\sexpone}{\sexpone}.
\]
\end{linenomath}

In the rest of this section, we assume a mapping from memory values to their
respective symbolic expression, and let the context discriminate the former from
the latter.

In Section~\ref{sec:infrules} we shall simplify symbolic expressions. For this,
we need a domain structure. Let us denote $\sexpsetinf = \sexpset \cup \{
\neginf; \posinf \}$ and $\zinf = \mathbb{Z} \cup \{ \neginf; \posinf \}$. We
define a \emph{valuation of a symbolic expression} $E$ every map
$\mathcal{V}(E)$, from $\sexpsetinf$ to $\zinf$, obtained by substituting every
\C variable in $E$ with a distinct integer, the symbol $\star$ with a natural
number strictly greater than 1 as a multiplicative coefficient, and interpreting
the operators $\{\kw{bop}, \kw{min}, \kw{max}\}$ as their respective functions
over $\zinf \times \zinf$. If we denote $\leq_\infty$ the standard ordering
relation on $\zinf$, then the preorder $\preccurlyeq$ on $\sexpsetinf$ is defined
as follows:
\begin{linenomath}
\[
E_1 \preccurlyeq E_2 \iff
\forall \mathcal{V},
\mathcal{V}(E_1) \leq_\infty \mathcal{V}(E_2).
\]
\end{linenomath}

The partial order $\preceq$ over $\sexpsetinf$ is therefore the one induced from
$\preccurlyeq$ by merging in the same equivalence class elements $x$ and $y$ of
$\sexpsetinf$ such that $x \preccurlyeq y$ and $y \preccurlyeq x$. As an example,
the elements $0$ and $\kw{min}(0,0)$ are equivalent.


\paragraph{Lattice of Symbolic Expression Ranges} A \emph{symbolic range}
$\srangeone$ is a pair of symbolic expressions $E_1$ and $E_2$, denoted
$\srange{\sexpone_1}{\sexpone_2}$. Otherwise said, a symbolic range is an
interval with no guarantee that $E_1 \preceq E_2$. We denote $\srangeset$ the
set of symbolic ranges extended with the empty range $\emptyset$ and
$\sqsubseteq$ its partial ordering which is the usual partial order over
(possibly empty) ranges. Any symbolic range $\srange{\sexpone_1}{\sexpone_2}$
such that $E_2 \prec E_1$ is therefore equivalent to $\emptyset$.
Consequently $(\srangeset, \sqsubseteq)$ is a domain. Its infimum is $\emptyset$
while its supremum is $\srange{-\infty}{+\infty}$. We denote $\sqcup$ and
$\sqcap$ its join and meet operators, respectively. It is worth noting that,
given $(E_i)_{1 \leq i \leq 4}$ four symbolic expressions, the following
equations hold:
\begin{linenomath}
\begin{align*}
\srange{E_1}{E_2} \sqcup \srange{E_3}{E_4}
&= \srange{\semin{E_1}{E_3}}{\semax{E_2}{E_4}}\\
\srange{E_1}{E_2} \sqcap \srange{E_3}{E_4}
&= \srange{\semax{E_1}{E_3}}{\semin{E_2}{E_4}}.
\end{align*}
\end{linenomath}

In words, \kw{min} and \kw{max} are compliant with our
ordering relations. In Section~\ref{sec:infrules}, when simplifying literals,
they will be introduced as soon as incomparable formul{\ae} will be associated
to the same left-value, resulting into an unsimplifiable constraint. Also, it is
worth noting that $\sqcup$ and $\sqcap$ are, in general, not statically
computable operators. To solve this practical issue, when these are not
computable on some symbolic expressions, \cfp relies on the above equations in
order to delay their evaluations at runtime. Eventually, the code generator will
convert them into conditionals.
\paragraph{State Constraints as Symbolic Ranges with Runtime Checks}
Symbolic ranges capture most minimal requirements over the \C left-values of a
function precondition: for integer typed left-values, a symbolic range
represents the integer variation domain, while for pointer typed left-values, it
represents a region of valid offsets. They are commonly used in abstract
interpreters for range~\cite{Cousot,Logozzo} and region
analysis~\cite{Na14,Rugina}, respectively.

However, some predicate literals cannot be simplified into symbolic ranges,
requiring their encoding as \emph{runtime checks}, that is, to be verified at
runtime by means of
conditionals. 
We denote $\test{\termone_1~\kw{cop}~\termone_2}$ 
a runtime check between two terms $\termone_1$ and $\termone_2$. We then call
\emph{state constraint} any pair $\cstrone=\srangeone\oplus\rtestone$ given by a
symbolic range $\srangeone$ and a set $\rtestone$ of runtime checks. We denote
$\pi_1(\cstrone)$ (resp. $\pi_2(\cstrone)$) the first (resp. the second)
projection of $\cstrone$, that is, $\srangeone$ (resp. $\rtestone$).

\subsection{Inferring State Constraints}
\label{sec:infrules}

We now formalize our solution for simplifying a positive literal into a set of
state constraints as a system of inference rules. Negative literals, as well as
conjunctive clauses, are handled separately at the end of the
section. 


\paragraph{Simplification Judgments}
Simplification rules are given over judgments of the form
\begin{linenomath}
$$
\predtocstr{\mem}{\predone}{\mem'},
$$
\end{linenomath}
where $\predone$ is a predicate literal, and $\mem$, $\mem'$ are maps from
left-values to state constraints. Each judgment associates a set of state
constraints $\mem$ and a literal $\predone$ with the result of simplifying
$\predone$ with respect to the left-values appearing in it, that is, an updated
map $\mem'$ equal to $\mem$ but for the state constraints on these latter.
Figures~\ref{fig:predicates2constraints} shows the formalization of the main
literal simplifications. This system does not assume the consistency of the
precondition: if this is inconsistent, no rule applies and the simplification
process fails.

\begin{figure*}[htbp]
\centering
\begin{subfigure}[h]{\textwidth}
  \[
  \inferrule[Idempotence]{
    \lvalone\in\mem}
  {\predtocstr{\mem}{\defpred{\lvalone}}{\mem}}
  \qquad\quad
  \inferrule[Variable]{
    \cvarone \not\in \mem\\
    \type{\cvarone}=\kappa\\
    \mem'=\memadd{\mem}{\cvarone}{\neutral{\kappa}}}
    {\predtocstr{\mem}{\defpred{\cvarone}}{\mem'}}
    \]
    \[
    \inferrule[Dereference]{
      \memacstwo{\memone}\not\in\mem\qquad
      \predtocstr{\mem}{\defpred{\memone}}{\mem'}\\
      \type{\memacstwo{\memone}}=\kappa \\
      \mem''=\memadd{\mem'}{\memacstwo{\memone}}{\neutral{\kappa}}}
      {\predtocstr{\mem}{\defpred{\memacstwo{\memone}}}{\mem''}}
      \]
      \[
      \inferrule[Range-1]{
        \proj{1}{\mem\left(\lvalone\right)}\not=\ival{\memone}{\memone}\\
        \predtocstr{\mem}
        {\defpred{\lvalone}\wedge 0\leq \termone_1 \leq \termone_2}{\mem'}\\
        \mem''=\memupone{\mem'}{\lvalone} {\proj{1}{\mem'\left(\lvalone\right)}
          \sqcup \ival{0}{\termone_2}}}
      {\predtocstr{\mem}{\defpred{\lvalone~\kw{++}\range{\termone_1}{\termone_2}}}{\mem''}}
        \]
      \[
      \inferrule[Range-2]{
        \proj{1}{\mem\left(\lvalone\right)}=\ival{\memone}{\memone}\\
        \tbase{\memone}=\lvaltwo\qquad
        \toffset{\memone}=\termone_3\\
        \predtocstr{\mem}{\defpred{\lvaltwo~\kw{++}\range{\semin{\termone_1}{\termone_3}}
            {\semax{\termone_2}{\termone_3}}}}{\mem'}}
        {\predtocstr{\mem}{\defpred{\lvalone~\kw{++}\range{\termone_1}{\termone_2}}}{\mem'}}
        \]
        \caption{Simplification of literal \kw{defined}.}
        \label{fig:defined}
      \end{subfigure}
      \hfill

      \begin{subfigure}[h]{\textwidth}
        \[
        \inferrule[Cmp-1]{
          \lvalone\in\{\termone_1,\termone_2\}\\
          \termone_1~\cmp~\termone_2 \rightsquigarrow \lvalone~\cmp~\termone_3 \\
          \predtocstr{\mem}
          {\defpred{\lvalone}\wedge\bigwedge_{\lvaltwo\in\termone_{3}}\defpred{\lvaltwo}}{\mem'}\\
          \mem''=\memupone{\mem'}{\lvalone}{\proj{1}{\mem'\left(\lvalone\right)} \sqcap
            \inter{\cmp}{\termone_3}}}
          {\predtocstr{\mem}{\termone_1~\cmp~\termone_2}{\mem''}}
          \]
          \[
          \inferrule[Cmp-2]{
            \predtocstr{\mem}{\bigwedge_{\lvalone\in\{\termone_1,\termone_2\}}\defpred{\lvalone}}{\mem'}\\
            \lvalone\in\{\termone_1,\termone_2\}\\
            \mem''=\memupone{\mem'}{\lvalone}
            {\proj{2}{\mem'\left(\lvalone\right)} \cup \test{\termone_1~\cmp~\termone_2}}}
          {\predtocstr{\mem}{\termone_1~\cmp~\termone_2}{\mem''}}
          \]
          \[
          \inferrule[Memory-Eq]{
            i,j\in\{1,2\}\wedge i\not= j\\
            \tbase{\memone_{\{i,j\}}}=\lvalone_{\{i,j\}}\\
            \toffset{\memone_{\{i,j\}}}=\termone_{\{i,j\}}\\
            \termone_3=\termone_j~\kw{+}~(-\termone_i)\\
            \memone'=\lvalone_j~\kw{++}\range{\termone_3}{\termone_3}\\
            \predtocstr{\mem}{\defpred{\lvalone_i}\wedge
              \defpred{\memone'}}{\mem'}\\
            \proj{1}{\mem'\left(\lvalone_i\right)} \sqsubseteq
            \proj{1}{\mem'\left(\lvalone_j\right)}\\
            \mem''=\memupone{\mem'}{\lvalone_i}
            {\ival{\memone'}{\memone'}}}
          {\predtocstr{\mem}{\memone_1 \equiv \memone_2}{\mem''}}
          \]
          \caption{Simplification of term comparison and memory equality literals.}
          \label{fig:comparisons}
      \end{subfigure}
\hfill
\vspace{.2cm}
\begin{subfigure}[h]{\textwidth}
\[
\inferrule[Not-Defined]
{\memone\not\in\mem}
{\predtocstr{\mem}{\neg\,\defpred{\memone}}{\mem}}
\]

\[
\inferrule[Memory-Neq]{
\predtocstr{\mem}{\defpred{\memone_1}\wedge\defpred{\memone_2}}{\mem'}\\
i,j\in\{1,2\}\wedge i\not= j\\
\tbase{\memone_{\{i,j\}}}=\lvalone_{\{i,j\}}\\
\ival{\lvalone_i}{\lvalone_i}\not\sqsubseteq\proj{1}{\mem'\left(\lvalone_j\right)}\qquad
\ival{\lvalone_j}{\lvalone_j}\not\sqsubseteq\proj{1}{\mem'\left(\lvalone_i\right)}
}
{\predtocstr{\mem}{\memone_1 \not\equiv \memone_2}{\mem'}}
\]
\caption{Simplification of negative literals.}
\label{fig:negatedpredicates}
\end{subfigure}

\caption{Simplification of literals into state constraints.}
\label{fig:predicates2constraints}
\end{figure*}

\paragraph{Predicates \textnormal{\kw{defined}}}
Figure~\ref{fig:defined} provides the simplification rules for literal
\kw{defined}. Rules \textsc{Variable} and \textsc{Dereference} enforce the
initialization of a left-value $\lvalone$ in terms of the symbolic range
$\neutral{\kappa}$. This latter is respectively defined as $\emptyset$, for
$\kappa$ a pointer type, and $\srange{\neginf}{\posinf}$, for $\kappa$ integer
type. These are quite common initial approximations when inferring variation
domains of either memory or integer values.

Rules \textsc{Range-1} and \textsc{Range-2} enforce the validity of a memory
region determined by the displacement range
$\lvalone~\kw{++}\range{\termone_1}{\termone_2}$. The first premise of these
rules established whether $\lvalone$ is already enforced in $\mem$ to be an
alias of a memory value $\memone$, as indicated by the singleton range
$\ival{\memone}{\memone}$. If not, rule \textsc{Range-1} first enforces the
initialization of $\lvalone$ and the soundness of the displacement bound
determined by $\termone_1$ and $\termone_2$, and then it updates the region of
valid offsets pointed to by $\lvalone$ to include the range
$\ival{0}{\termone_2}$. In practice, predicates $0\leq \termone_1\leq
\termone_2$ are added only if not statically provable. Moreover, note that we do
not consider $\termone_1$ as the lower bound of the symbolic range, because \C
memory regions must start at index $0$. Rule \textsc{Range-2} handles the case
of $\lvalone$ alias of $\memone$ in $\mem$ by enforcing the validity of the
memory region determined by $\memone$ to take into account the displacement
range $\range{\termone_1}{\termone_2}$. In particular, since single
displacements only may appear in memory equality predicates (\cf rule
\textsc{Memory-Eq}), $\memone$ is of the form
$\lvalone'~\kw{++}\range{\termone_3}{\termone_3}$, and the validity of the alias
$\lvalone$ within the range $\range{\termone_1}{\termone_2}$ is obtained by
requiring the validity of the displacement range
$\lvaltwo~\kw{++}\range{\semin{\termone_1}{\termone_3}}{\semax{\termone_2}{\termone_3}}$.

Rule \textsc{Idempotence} is provided only to allow the inference process to
progress.

\paragraph{Term comparison predicates}
Rules in Figure~\ref{fig:comparisons} formalize the simplification of integer
term comparison and memory equality predicates. The first two are actually rule
schema, as \textsc{Cmp-1} and \textsc{Cmp-2} describe term comparison
simplifications over the integer comparison operators
$\{\equiv,\,\leq,\,\geq\}$. (Strict operators are treated in terms of non-strict
ones.) Let us detail rule \textsc{Cmp-1} with respect to a generic operator
$\cmp$. The rule applies whenever $\termone_1~\cmp~\termone_2$ can be rewritten
by means of classic integer arithmetic transformations as
$\lvalone~\cmp~\termone_3$, that is, as a left-value
in relation $\cmp$ with an integer term
$\termone_3$. If so, \textsc{Cmp-1} reduces the symbolic range of $\lvalone$
with respect to the one given by $\inter{\cmp}{\termone_3}$. This latter
function takes a comparison operator $\cmp$ and an integer term $\termone$ as
arguments, and returns as result the symbolic range $\ival{\termone}{\termone}$
when $\cmp$ is $\equiv$, $\ival{\neginf}{\termone}$
(resp. $\ival{\termone}{\posinf}$) when $\cmp$ is $\leq$ (resp.
$\geq$). 
Since both $\lvalone$ and $\termone_3$ are integer typed terms, there is no
aliasing issue here. Rule \textsc{Cmp-2} can always be applied, although we
normally consider it when \textsc{Cmp-1} cannot. In that case, rule
\textsc{Cmp-2} conservatively enforces the validity of the term comparison by
means of a runtime check.

\paragraph{Aliasing}
Rule \textsc{Memory-Eq} handles aliasing between two pointers with single
displacement $\memone_1$ and $\memone_2$. Assuming both of the form
$\lvalone_{\{i,j\}}~\kw{++}~\termone_{\{i,j\}}$, with distinct $i,j\in\{1,2\}$,
a pointer $\memone'$ is first defined as $\lvalone_j$ with single displacement
$\termone_3$, this latter determined by summing the offsets $-\termone_i$ and
$\termone_j$ together. Such a pointer is then enforced to be defined, and in the
case that the actual region pointed by $\lvalone_j$ is established to be larger
then the one pointed by $\lvalone_i$, then $\lvalone_i$ is considered an alias
of $\memone'$. Although rather conservative, due to the fact that $\sqsubseteq$
is not statically computable in general, the second to last premise is important
for ensuring soundness.

\paragraph{Negative literals}

Figure~\ref{fig:negatedpredicates} shows the rules for negative literals. These
rules do not simplify literals into state constraints, but rather ensure
precondition consistency. For instance, $\neg\defpred{\cvarone}\wedge
\cvarone==0$ is inconsistent as $\cvarone$ should be defined with value $0$ and
undefined at the same time. In such a case, the system must prevent code
generation.


Rule \textsc{Not-Defined} just checks that the memory value $\memone$ does not
appear in the map $\mem$, which suffices to ensure that $\memone$ is not yet
defined.

Rule \textsc{Memory-Neq} applies under the hypothesis that both pointers $M_1$
and $M_2$
determine different memory regions. In particular, the two are not aliases
whenever each base address of one pointer does not overlap with the memory
region of the other.

\paragraph{Conjunctive Clauses}
$\bigwedge_i \predone_i$, on either positive or negative literals $\predone_i$,
are handled sequentially through the following \textsc{And} rule. Given the
definition of \textsc{Memory-Neq} and \textsc{Not-Defined}, it assumes that
negative literals are treated only after the positive ones, by exhaustively
applying rule \textsc{Memory-Neq} first, and rule \textsc{Not-Defined}
afterwards.

{\small 
  \[
  \inferrule[And]{\predtocstr{\mem_0}{\predone_1}{\mem_1}\\
    \predtocstr{\mem_1}{\predone_2}{\mem_2}\\
  \cdots\\
    \predtocstr{\mem_{n-1}}{\predone_n}{\mem_n}}
  {\predtocstr{\mem_0}{\bigwedge_i \predone_i}{\mem_n}}
  \]
}

\paragraph{Dependency Graph on Memory Values}

On a conjunctive clause, the system of inference rules in
Figure~\ref{fig:predicates2constraints} not only generates a map $\mem$, but it
also computes a dependency graph $\mathcal{G}$ on memory values. (Considering
only the formalization of this section, the memory values of the graph are
actually left-values only. However, when considering separately the \acsl
predicates \lstinline+\initialized+ and \lstinline+\valid+ instead of
\kw{defined}, this is not true anymore.) This graph is necessary for ensuring,
first, the soundness of the rule system with respect to mutual dependency on
left-values in $\mem$, and, consequently, for the correct ordering of left-value
initializations when generating \C code (\cf Section~\ref{sec:cstr_to_code}).

Generally speaking, each time a rule that needs inference is used in a state
constraint derivation for some left-value $\lvalone$ (\eg{}
\textsc{Dereference}, \textsc{Range-1}, \textsc{Cmp-1}, \etc{}), edges from
$\lvalone$ to every other left-value involved in some premise are added to the
dependency graph $\mathcal{G}$. Such derivation fails as soon as this latter
operation makes the graph $\mathcal{G}$ cyclic.

\begin{figure}[htb]
\vspace{-6mm}
\centering
\includegraphics[scale=0.4]{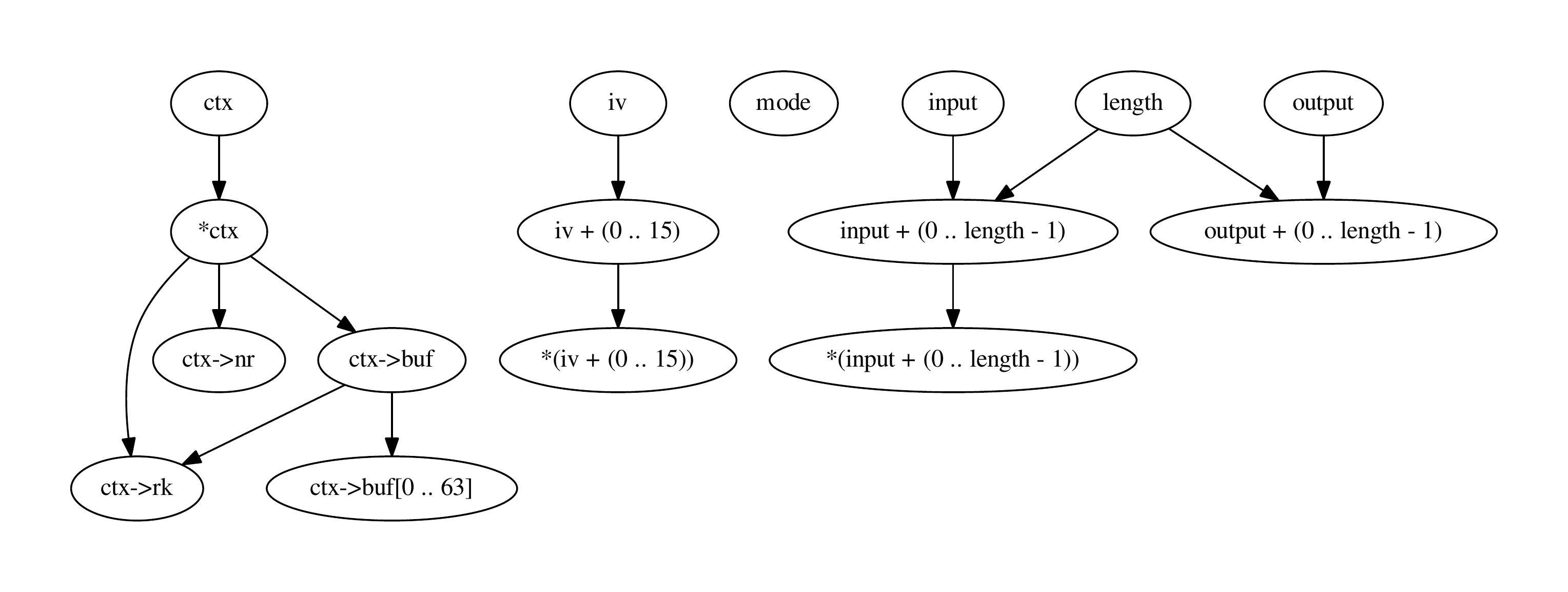}
\caption{Dependency graph for the \kw{aes\_crypt\_cbc} preconditions generated
by \cfp.}
\vspace{-3mm}
\label{fig:depgraph}
\vspace{-2mm} 
\end{figure}

\paragraph{Example}
When applying the inference system on our example in Figure~\ref{fig:contract},
the final map associates the integer \texttt{length} to $\srange{16}{16672}
\oplus \{ \test{\texttt{length} \% 16 \equiv 0} \}$ and the array \texttt{input}
to $\srange{0}{\texttt{length}-1} \oplus \emptyset$, along with the dependency
graph in Figure~\ref{fig:depgraph}.\\

The system of inference rule in Figure~\ref{fig:predicates2constraints} is
sound: given a conjunctive clause $\predone$, the simplification procedure on
$\predone$ always terminates, either with $\mem$ or it fails. In the former
case, for each left-value $\lvalone$ in $\predone$, state constraints in $\mem$
satisfy respective literals in $\predone$ (that we denote as $\mem \models
\predone$).

\begin{theorem}
\label{thm:soundness}
  For all conjunctive clause $\predone$, either $\emptyset \vdash \predone
  \Rightarrow \mem$ and $\mem \models \predone$, or it fails.
\end{theorem}

\vspace{-4mm} 
\section{Generating \C Code from State Constraints}
\label{sec:cstr_to_code}
\vspace{-2mm} 

This section presents the general scheme for implementing preconditions, through
state constraints, in a \C language enriched with one primitive function for
handling ranges. 
In practice, such primitive is meant to be analyzer-specific so as to
characterize state constraints as precisely as possible. As an example, we
report on the case of our tool \cfp. However, for the sake of conciseness, we do
neither detail nor formalize the code generation
scheme. 
We nevertheless believe that the provided explanation should be enough to both
understand and implement such a system in a similar setting.

\vspace{-2mm} 
\paragraph{Generating Code from a Conjunctive Clause}

Consider a conjunctive clause $\mathcal{C}$ and the pair ($\mem$,
$\mathcal{G})$, respectively given by the map of state constraints and the
dependency graph of $\mathcal{C}$, inferred by the system of rules in
Figure~\ref{fig:predicates2constraints}. We shall show the general case of
disjunctive normal forms $\bigvee_{i=1}^n \mathcal{C}_i$ later on.

To generate semantically correct \C code, we topologically iterate over the
left-values of $\mathcal{G}$ so as to follow the dependency ordering. For every
visited left-value $\lvalone$, we consider its associated state constraint
$\cstrone=\srangeone\oplus\rtestone$ in $\mem$. Then, the symbolic range
$\srangeone$ is handled by generating statements that initialize $\lvalone$. For
most constructs, these statements are actually a single assignment, although a
loop over an assignment may be sometimes needed (\eg{} when initializing a range
of array cells). In particular, initializations of left-values $\lvalone$ to
symbolic ranges $\srange{\termone_1}{\termone_2}$ are implemented by means of
the primitive function \kw{make\_range}$(\kappa, \termone_1, \termone_2)$, where
$\kappa$ is integer or pointer type. In practice, this function must be provided
by the analyzer for which the context is generated, so that, when executed
symbolically, the analyzer's abstract state will associate abstract values
$\srange{\termone_1}{\termone_2}$ to respective left-values
$\lvalone$. 
Finally, conditionals are generated to initialize left-values with symbolic
expressions involving \kw{min} and \kw{max}.

Once $L$ has been initialized, the rest of the code is guarded by conditionals
generated from runtime checks in $X$. To resume, the generation scheme for $L$
is the following:
\begin{lstlisting}[deletekeywords={for}]
  /* initialization of L from R through assignments */
  if (/* runtime checks from X */) {
    /* code for initializing the next left-values */ ...; } }
\end{lstlisting}

After the initialization of the last left-value, the function under
consideration (in our running example, the function \texttt{aes\_crypt\_cbc}) is
called with the required arguments.

\vspace{-2mm} 
\paragraph{Handling Disjunctions}

We rewrite preconditions into disjunctive normal form $\bigvee_{i=1}^n
\mathcal{C}_i$ as a preliminary step. Then we process each disjunct
$\mathcal{C}_i$ independently by applying the inference system in
Figure~\ref{fig:predicates2constraints} and the code generation scheme
previously described.

We now describe the code generation scheme of such a precondition
$\bigvee_{i=1}^n \mathcal{C}_i$ given the code fragments for each and every of
its disjunct $\mathcal{C}_i$. If $n=1$, then the code fragment of
$\mathcal{C}_1$ is directly generated. Otherwise, an additional variable
\texttt{cfp\_disjunction} is generated and initialized to the interval
$\srange{1}{n}$. Then, a \texttt{switch} construct (or a conditional if $n=2$)
is generated, where each case contains the fragment $B_i$ respective to
$\mathcal{C}_i$. To resume, the context is generated as a function including the following
code pattern:

\begin{lstlisting}[escapeinside={(*}{*)}]
  cfp_disjunction = make_range((*$\iota$*), 1, n);
  switch (cfp_disjunction) {
    case 1: { B_1; break; }
    case 2: { B_2; break; }
    ...
    case n: { B_n; break; }
  }
\end{lstlisting}

\vspace{-4mm} 
\newcommand{\tautau}{\ensuremath{\tau}}
\paragraph{Primitives in \cfp}
Our tool \cfp follows the generation scheme just described. It implements
\kw{make\_range} in terms of the \framac built-ins
\texttt{Frama\_C\_\tautau\_interval}, with $\tautau$ a \C integral type, and
\texttt{Frama\_C\_make\_unknown} to handle symbolic ranges for integers and
pointers, respectively. These built-ins are properly supported by the two
abstract interpretation-based value analysis tools \Eva~\cite{blazy17vmcai} and
\tisanalyzer~\cite{cuoq17jfla}.


\vspace{-3mm} 
\section{Implementation and Evaluation}
\label{sec:evaluation}

We have implemented our context generation mechanism as a \framac plug-in,
called \cfp for \emph{Context from Preconditions}, written in approximately 3500
lines of \caml. (Although \framac is open source, \cfp is not, due to current
contractual obligations.) \cfp has been successfully used by the company \tis
for its verification kit~\cite{polarssl-report} of the \polarssl library, an
open source implementation of the SSL/TLS protocol.

We now evaluate our approach, and in particular \cfp, in terms of some quite
natural properties, that is, \emph{usefulness}, \emph{efficiency}, and
\emph{quality} of the generated contexts.

This work provides a first formal answer to a practical and recurring problem
when analyzing single functions. Indeed, the \acsl subset considered is
expressive enough for most real-world \C programs. Most importantly, \cfp
enables any tool to support a compelling fragment of \acsl at the minor expense
of implementing two \framac built-ins, particularly so if compared to the
implementation of a native support (if ever possible). Finally, \cfp has proved
useful in an operational industrial setting in revealing some mistakes in
contexts previously written by hand by expert verification engineers. Although
we cannot disclose precise data about these latter, \cfp revealed, most notably,
overlooked cases in disjunctions and led to fix incomplete specifications.

\cfp is able to efficiently handle rather complex \acsl preconditions: the
generation of real-world contexts (\emph{e.g.} the one of
Figure~\ref{fig:generate}) is usually instantaneous. Although the disjunctive
normal form can be exponentially larger
than the original precondition formula, such transformation is used in
practice~\cite{pugh92,kuncak10pldi} and leads to better code in terms of
readability and tractability by the verification tools. This approach is further
justified by the fact that, in practice, just a small number of disjuncts are
typically used in manually-written \acsl specifications.

Our approach allows to generate contexts which are reasonably readable and
follows code patterns that experts of the \framac framework use to manually
write. In particular, when handling disjunctions, \cfp factorizes the generated
code for a particular left-value as soon as the rule system infers the very same
solution in each conjunctive clause. For instance, in our running example, only
the initialization of the variable \texttt{mode} depends on the disjunction
\texttt{mode == 0 || mode == 1}. Hence all the other left-values are initialized
before considering \texttt{cfp\_disjunction} (\cf{} Figure~\ref{fig:generate}).

We conclude by briefly discussing some current limitations. Our \acsl fragment
considers quantifier free predicate formul{\ae}, and no coercion constructs are
allowed. Support for casts among integer left-values should be easy to add,
whereas treating memory addresses as integers is notoriously difficult. We leave
these for future work.

\vspace{-3mm} 
\section{Related Work}
\label{sec:related_work}
\vspace{-1mm} 

Similarly to our approach, program
synthesis~\cite{kuncak10pldi,lezama07,polik16} automatically provides program
fragments from formal specifications. However, the two approaches have different
purposes. Once executed either symbolically or concretely, a synthesized program
provides \emph{one} computational state that satisfies the specification, while
a context must characterize \emph{all} such states. In particular, not only
every state must satisfy the specification but, conversely, this set of states
must contain every such possible one.

In software testing, contexts are useful for concentrating the testing effort on
particular inputs. Most test input generation tools, like \tool{CUTE}
\cite{sen05} and \tool{PathCrawler} \cite{botella09,delkosm13}, allow to express
contexts as functions which, however, the user must manually write. Some others,
like \tool{Pex}~\cite{barnett09}, directly compile formal preconditions for
runtime checking.

The tool \tool{STADY}~\cite{petiot14} shares some elements of our
approach. It instruments \C
functions with additional code for ensuring pre- and postconditions compliance,
allowing monitoring and test generation. However, the tool performs a simple
\acsl-to-\C translation, it does neither take into account dependencies among \C
left-values, nor it inferences their domain of definition.

\vspace{-3mm} 
\section{Conclusion}
\label{sec:conclusion}
\vspace{-2mm} 

This paper has presented a novel technique to automatically generate an analysis
context from a formal precondition of a \C function. The core of the system has
been formalized, while we provide enough details about code generation to allow
similar systems to be implemented. Future work includes the formalization of
code generation as well as statements and proofs of the fundamental properties
of the system as a whole. A running example from the real world has also
illustrated our presentation. The whole system is implemented in the \framac
plug-in \cfp. It generates code as close as possible to human-written code. It
is used in an operational industrial setting and already revealed some mistakes
in contexts previously written by hand by expert verification engineers.

\vspace{-2mm}

\section*{Acknowledgments}
\vspace{-1mm} 
Part of the research work leading to these results has received funding for the
S3P project from French DGE and BPIFrance. The authors thank \tis for the
support and, in particular, Pascal Cuoq, Benjamin Monate and Anne Pacalet for
providing the initial specification, test cases and insightful comments. Thanks
to the anonymous reviewers for many useful suggestions and advice.

\vspace{-2mm}




\bibliographystyle{abbrv}
\bibliography{main}


%

\end{document}